\documentclass[runningheads]{llncs}
\usepackage{graphicx}
\usepackage{xcolor}
\usepackage{amsmath,amssymb,amsfonts}
\usepackage{textcomp}
\usepackage{crayola}
\usepackage{listings}
\usepackage{amsmath}
\usepackage{algorithm}
\usepackage{listings}
\usepackage{wasysym}
\usepackage{marvosym}
\usepackage{graphicx}
\usepackage{tikz}
\usepackage{framed}
\usepackage{tikz-uml}
\usepackage{alltt}
\usepackage{mathpartir}
\usepackage{algorithm}
\usepackage[noend]{algpseudocode}
\usepackage{tabularx}
\usepackage[belowskip=-15pt,aboveskip=0pt]{caption}
\usepackage{soul, xcolor}
\usepackage{xfrac}
\usetikzlibrary{positioning,shapes,arrows,backgrounds,fit, calc, shadows}
\usepackage[skins]{tcolorbox}

\newcommand{\btl}{\blacktriangleleft}
\newcommand{\btr}{\blacktriangleright}

\algrenewcommand\algorithmicindent{1.0em}%

\newtcolorbox{myframe}[2][]{%
  enhanced,colback=white,colframe=black,coltitle=black,
  sharp corners,boxrule=0.4pt,
  fonttitle=\itshape,
  attach boxed title to top left={yshift=-0.3\baselineskip-0.4pt,xshift=2mm},
  boxed title style={tile,size=minimal,left=0.5mm,right=0.5mm,
    colback=white,before upper=\strut},
  title=#2,#1
}

\begin{document}
\title{StaBL - State Based Language for Specification of Web Applications}
%
%\titlerunning{Abbreviated paper title}
% If the paper title is too long for the running head, you can set
% an abbreviated paper title here
%
\author{Karthika Venkatesan\inst{1,2} \and
Sujit Kumar Chakrabarti\inst{1}}
\authorrunning{Karthika Venkatesan, Sujit Kumar Chakrabarti}
% First names are abbreviated in the running head.
% If there are more than two authors, 'et al.' is used.
%
\institute{ International Institute of Information Technology, Bangalore \\
\email{karthika.venkatesan@iiitb.org,sujitkc@iiitb.ac.in}\\ \url{http://www.iiitb.ac.in}
 \and
C-DAC, Bangalore}
\maketitle              % typeset the header of the contribution
\lstset{
	language = Caml,
	basicstyle = \small\ttfamily,
	stringstyle = \color{black}\ttfamily,
	keywordstyle=\color{black}\bfseries,
	identifierstyle=\ttfamily,
	frameround=tttt,
	%numbers=left,
	showstringspaces=false
}

\lstdefinestyle{mycode}{
	language = Caml,
	basicstyle = \scriptsize\ttfamily,
	stringstyle = \color{black}\ttfamily,
	keywordstyle=\color{black}\bfseries,
	identifierstyle=\ttfamily,
	frameround=tttt,
	numbers=none,
	showstringspaces=false,
	escapeinside={(*@}{@*)}
}

\setstcolor{red}
\begin{abstract}
\textbf{[Context and motivation]} Usage of Formal Specification languages is scarce in web application development as compared to safety critical/hardware systems. \textbf{[Question/problem]} An apt formal specification language should provide the following features: Firstly, it should have well-defined semantics, so that specifications written in it can not be inherently ambiguous. Secondly, it should have tool support for automatic detection of specification bugs. Additionally, for domains like web development, it is important that specification formalisms build over familiar notations, as the benefits of learning highly mathematical notations in such domains are perceived to be low. \textbf{[Principal ideas/results]} This work presents a \textbf{Sta}te \textbf{B}ased \textbf{L}anguage inspired by Statecharts called StaBL for specification of web applications, and how StaBL can be used for writing such specifications. We also present modifications to the language w.r.t Statechart which facilitate writing modular and scalable specification. \textbf{[Contribution]} In particular, we present the feature of \emph{locally scoped variables} with \emph{inter-state data-flow}. We summarise our experience of developing specifications with StaBL, which shows that StaBL specifications, on the one hand, are able to capture most essential elements of the functional aspects of a web application while foregoing much of the verbosity of a regular programming language. 

\end{abstract}
\section{Introduction}
Statecharts\cite{HAREL1987231} is a visual notation for specification of complex systems. In particular, the `system' here is an object with a complex life-cycle going through a finite number of discrete \emph{states}. Statecharts\footnote{We use Statechart and Statecharts to refer to the notation while statechart(s) refers to specific models developed in this notation.} notation has enjoyed a wide acceptance among software engineering practitioners useful in specifying requirements and design in a semi-formal manner. It is a part of the UML suite for modelling object oriented systems, and has numerous popular commercial implementations, e.g. IBM Rational Rose, Matlab Stateflow and Rhapsody. There has been a lot of work done in defining formal semantics for Statecharts\cite{10.1007/978-3-540-24721-0_17} and developing verification technology around it. However, a lot of this work focuses on safety critical systems (e.g. embedded systems and cyber physical systems). The so called non safety critical domains of software development (e.g. web-, desktop- and mobile applications) have grown complex and critical enough to warrant support for formal specification and verification. General adoption of existing formal specification languages like Z\cite{Potter:1996:IFS:547639}, B-method\cite{Alagar2011}, TLA+\cite{1033032} and Alloy\cite{Jackson:2002:ALO:505145.505149} in non safety critical domains is hampered due to the widely held perception that such languages are highly mathematical and hard to use. Therefore, need is felt to develop formal verification technology for non safety critical domains built on more `accessible' (as in `familiar') notations like Statecharts.

Earlier attempts to use Statecharts for specifying web applications have focused on modelling specific aspects of web applications, e.g. navigation\cite{884689,DEUTSCH2007442,748917}, web-service composition\cite{1240303} etc. To the best of our knowledge, there has been limited or no investigation to use Statecharts as a general purpose specification language for web applications -- revisiting its features to adapt it to suit such applications -- with complete integration with formal verification.

In this paper, we introduce \textbf{StaBL}, a \textbf{Sta}te \textbf{B}ased \textbf{L}anguage for specifying web applications. StaBL takes inspiration from well-known imperative languages like C and Python (mutable variables, instructions, lexical scoping etc.), but borrows significantly some important aspects from Statecharts: \emph{viz.}, states, transitions, hierarchy and action language. We present our language both informally through examples as well as through formal description of some salient aspects of its semantics. Our examples also demonstrate how formal specifications for web applications can be written simply and intuitively using StaBL. The specification engineer uses her familiar programmatic constructs with little or no need to learn a very different looking language. The nitty-gritty implementation details of the system being specified can be abstracted away while maintaining the formal nature of the specification. This paper we present StaBL language, with its salient syntax, semantic and usage aspects

The paper is structured as follows: Section~\ref{s:stabl} introduces StaBL, our variant of Statechart, and illustrates how to specify a web application using it. Sections~\ref{s:ld} and \ref{s:sem} present the details of design aspects of StaBL in particular of its syntax and type system. Section~\ref{s:rw} survey related works and Section~\ref{s:conc} concludes the paper.

\section{Specification using StaBL} \label{s:stabl}
In this section, we present an brief overview of StaBL (\textbf{Sta}te-\textbf{B}ased \textbf{L}anguage). StaBL is a language with constructs supporting programming using primarily imperative style. These imperative elements make its syntax approachable to a developer. On the other hand, StaBL's main objective is to write functional specifications for web applications. This is facilitated by the fact that StaBL code is organised within a hierarchical state machine very much like a statechart. The state-based structure also makes it easy to visualise a StaBL program in a pictorial form akin to Statecharts.

We introduce StaBL's basic features through an example.
\subsection{Example}

\tikzstyle{init} = [draw=black, thick, fill=black, circle]
\tikzstyle{tr} = [rectangle,draw=black, very thin, dashed,align=left,font=\scriptsize, text ragged, minimum height=1em, minimum width=2em, inner sep=1pt, outer sep=3pt]
\tikzstyle{trl} = [->, thick, color=Purple, text=black]
\tikzset{obj/.style = {rectangle split, rounded corners,
                      rectangle split parts=2, thick,draw=black, top
                      color=Dandelion!10,bottom color=Dandelion!20, align=center}}

\tikzset{cobj/.style = {rectangle split, rounded corners,
                      rectangle split parts=3, thick,draw=black, top
                      color=white,bottom color=Dandelion!10, align=center}}

\tikzset{sn/.style = {rectangle, draw=gray, fill=blue!10, thick, align=center, rounded corners}}
\tikzset{sdn/.style = {ellipse, draw=gray, fill=red!10, thick, align=center}}
\tikzset{st/.style = {-latex, draw=brown, thick}}

\newcommand\Mstrut[1]{\rule{0pt}{#1cm}}
\begin{figure*}

\begin{center}
\resizebox{0.9\textwidth}{!}{
\begin{tikzpicture}

\node[cobj, minimum width = 14cm, minimum height = 30cm](student) at (0, 0) {
  \nodepart{one} \textbf{Student}
  \nodepart{two} \lstinline[mathescape=true]@  students : map$\btl$int, string$\btr$@ \\
      \lstinline[mathescape=true]@rooms : map$\btl$int, int$\btr$@
  \nodepart{three} {\Mstrut{9.5}}
};

\node[cobj, above = 0.5cm of student.south, xshift=-0cm, minimum width = 13cm](loggedin) {
  \nodepart{one} \textbf{LoggedIn}
  \nodepart{two} \lstinline@loggedinUser : int@
  \nodepart{three} {\Mstrut{5}}
};

\node[obj](loggedout) [above = 1cm of loggedin] {
  \nodepart{one} \textbf{LoggedOut}
  \nodepart[font=\scriptsize]{two} \lstinline@user # : int@ \\ \lstinline@password # : string@
};

\node[init](in1) [left = of loggedout]{};

\draw[trl](loggedout) to node[tr, left]{$t_{login}$} (loggedin);
\draw[trl, bend right](loggedin) to node[tr, right]{$t_{logout}$} (loggedout.east);

\draw[trl](in1) to node[tr, above]{$t_{in1}$} (loggedout);

\node[obj, minimum width = 3cm](setroom) [above right = of loggedin.south west, xshift=-0.5cm, yshift=-0.5cm] {
  \nodepart{one}\textbf{SetRoom}
  \nodepart{two}\lstinline@room # : int@
};

\node[obj](viewdetails) [right = of setroom, xshift= 4cm, minimum width = 4cm] {
  \nodepart{one}\textbf{ViewDetails}
  \nodepart{two}\lstinline@student, room : int@
};

\node[obj](dashboard) [above right = of setroom, xshift=-0.5cm, yshift=1cm, minimum width = 5cm] {
  \nodepart{one}\textbf{Dashboard}
  \nodepart{two}
};

\node[init](in2) [left = of dashboard, yshift=0.5cm]{};

\draw[trl](dashboard) --node[tr, right]{$t_{setroom}$}(setroom);
\draw[trl](dashboard) --node[tr, left]{$t_{viewdetails}$}(viewdetails);
\draw[trl, bend left](setroom.north) to node[tr, above]{$t_{setroom\_done}$}(dashboard.west);
\draw[trl, bend right](viewdetails.north) to node[tr, near start, right]{$t_{viewdetails\_done}$}(dashboard.east);
\draw[trl, bend left](in2) to (dashboard.north west);
  \end{tikzpicture}
}

(a)
\end{center}

\begin{center}

\scalebox{1}{
\lstset{style=mycode}

\begin{tabular}{| l | p{10cm} |}
\hline
\textbf{Transition} & \textbf{Code} \\
\hline
\hline
$t_{in1}$ &
\begin{minipage}{0.5\textwidth}
\begin{tabular}{l @{\hspace{0.2cm}:\hspace{0.5cm}} l}
\textbf{action}  &  \\
\end{tabular}
\end{minipage}
\\
\hline
$t_{login}$ &
\begin{minipage}{1.0\textwidth}
\begin{tabular}{l @{\hspace{0.2cm}:\hspace{0.5cm}} l}
\textbf{trigger} & \lstinline@eLogin@ \\
\textbf{guard}   & \lstinline[mathescape=true]
@get_map$\btl$int, string$\btr$(students, Student.LoggedOut.user)@ \\
& \lstinline[mathescape=true]@  = Student.LoggedOut.password@
 \\
\textbf{action}  & \lstinline[mathescape=true]@Student.LoggedIn.loggedinUser $\gets$ Student.LoggedOut.user@ \\
\end{tabular}
\end{minipage}
\\
\hline
$t_{logout}$ &
\begin{minipage}{0.5\textwidth}
\begin{tabular}{l @{\hspace{0.2cm}:\hspace{0.5cm}} l}
\textbf{trigger} & \lstinline@eLogout@ \\
%\textbf{guard}   & \lstinline@true@ \\
\textbf{action}  &      \lstinline[mathescape=true]@Student.LoggedOut.user $\gets$ 0@ \\
      & \lstinline[mathescape=true]@Student.LoggedOut.password $\gets$ ""@
\\
\end{tabular}
\end{minipage}
\\
\hline
$t_{setroom}$ &
\begin{minipage}{0.5\textwidth}
\begin{tabular}{l @{\hspace{0.2cm}:\hspace{0.5cm}} l}
\textbf{trigger} & \lstinline@eSetRoom@ \\
%\textbf{guard}   & \lstinline@true@ \\
%\textbf{action}  &  \\
\end{tabular}
\end{minipage}
\\
\hline
$t_{viewdetails}$ & 
\begin{minipage}{0.5\textwidth}
\begin{tabular}{l @{\hspace{0.2cm}:\hspace{0.5cm}} l}
\textbf{trigger} & \lstinline@eViewDetails@ \\
%\textbf{guard}   & \lstinline@true@ \\
%\textbf{action}  &  \\
\end{tabular}
\end{minipage}
\\
\hline
$t_{setroom\_done}$ & 
\begin{minipage}{0.5\textwidth}
\begin{tabular}{l @{\hspace{0.2cm}:\hspace{0.5cm}} l}
\textbf{trigger} & \lstinline@eDoneSetting@ \\
%\textbf{guard}   & \lstinline@true@ \\
%\textbf{action}  &  \\
\end{tabular}
\end{minipage}
\\
\hline
$t_{viewdetails\_done}$ & 
\begin{minipage}{0.5\textwidth}
\begin{tabular}{l @{\hspace{0.2cm}:\hspace{0.5cm}}l}
\textbf{trigger} & \lstinline@eDoneViewing@ \\
%\textbf{guard}   & \lstinline@true@ \\
%\textbf{action}  &  \\
\end{tabular}
\end{minipage}
\\
\hline
\end{tabular}
}
\end{center}

\begin{center}
\lstset{style=mycode}

\begin{tabular}{| l | p{10cm} |}

\hline
\textbf{State} & \textbf{Code} \\
\hline
\hline
\textbf{Student} &
\begin{minipage}{0.5\textwidth}
\begin{tabular}{l @{\hspace{0.2cm}:\hspace{0.5cm}} l}
\textbf{entry} & \lstinline[mathescape=true]@students $\gets$ put_map$\btl$int, string$\btr$(students, 1, "p1")@ \\
 & \lstinline[mathescape=true]@students $\gets$ put_map$\btl$int, string$\btr$(students, 2, "p2")@\\
%\textbf{exit}   &  \\
\end{tabular}
\end{minipage}
\\
\hline
\textbf{SetRoom} &
\begin{minipage}{0.5\textwidth}
\begin{tabular}{l @{\hspace{0.2cm}:\hspace{0.5cm}} l}
%\textbf{entry} & \\
\textbf{exit}   & \lstinline[mathescape=true]@rooms $\gets$ put_map$\btl$int, int$\btr$(rooms, loggedinUser, room)@ \\
\end{tabular}
\end{minipage}
\\
\hline
\textbf{ViewDetails} &
\begin{minipage}{0.5\textwidth}
\begin{tabular}{l @{\hspace{0.2cm}:\hspace{0.5cm}} l}
\textbf{entry} &
\lstinline[mathescape=true]
@student $\gets$ loggedinUser;@ \\
& \lstinline[mathescape=true]
@room    $\gets$ get_map$\btl$int, int$\btr$(rooms, loggedinUser)@
\\
%\textbf{exit}   &  \\
\end{tabular}
\end{minipage}
\\

\hline
\end{tabular}

\vspace{0.5cm}

(b)

\end{center}
\caption{StaBL specification for simple website}
\label{f:st}
\end{figure*}

Fig.~\ref{f:st} shows an example of a specification written in StaBL.  We have shown the logical structure of the specification in fig.~\ref{f:st}(a) (in our current implementation, we use a linear concrete syntax akin to regular programming languages; but a pictorial view is more readable to an unaccustomed reader). The specification describes a simple web application named \textbf{Student} with two main states: \textbf{LoggedOut} and \textbf{LoggedIn}.  A transition $t$ gets enabled when the machine is the source state state of $t$, the event on its trigger takes place and its guard evaluates to true. Code can be inserted in two places: firstly, as entry/exit code inside states; secondly, as action code on transitions. For readability, we have listed this code separately in a tabular form in fig.~\ref{f:st}(b).
\textbf{LoggedIn} internally has sub-states \textbf{SetRoom} and \textbf{ViewDetails}. StaBL states are all OR states, which means that at any time the state machine is allowed to be in only one configuration involving a leaf state and all its ancestors. All these features are inspired by traditional Statecharts.

\section{Language Design} \label{s:ld}
Now, we briefly discuss some of the design choices that were considering while creating StaBL and present the features which are distinct from traditional Statecharts. These are primarily related to the scoping rules to suit the various specification requirements.

\subsection{Syntax and Usage} \label{s:syn}
\subsubsection{Local variables}

States may have local variables, e.g. in fig.~\ref{f:st}, \textbf{LoggedOut} has two local variables \lstinline@user@ and \lstinline@password@. A hash-sign (\lstinline@#@) against the variable name indicates that it is an input variable. Local variables are important to ensure a scalable development in the same manner as in the case of programming languages. Variable declarations can occur at two types of places: within states and within action code blocks (both in transition actions and in entry/exit blocks of states).

\subsubsection{Data flow}

Web applications are characterised by data-flow between views and pages. The precise mechanisms may vary from normal client side shared variables to message passing through the server. Hence, it is required to permit controlled access to local variables of states from outside it. We discuss this issue further on section~\ref{s:env} below.
\subsubsection{Types}
In its current state, StaBL provides basic types like \lstinline@int@, \lstinline@boolean@ and \lstinline@string@ with common set of operators, and generic types like \lstinline@map@, \lstinline@list@ and \lstinline@set@. There are also standard functions associated with these types (e.g. \lstinline@get_map@ and \lstinline@put_map@ -- all of which are generic -- for \lstinline@map@). Having generic containers allows defining a variety of container objects without sacrificing type safety. More details on types and its implementation is given in Section \ref{s:types}

\subsubsection{Imperative style}
Although, a language like Java or C++ is very well-suited as the action language in writing Statechart specifications, a purely object oriented style imposes syntactic overheads (e.g. class definitions) which are not desirable while specifying functionality at a high level. On the other hand, a language like C with no support for parametric polymorphism does not meet the basic standards of type safety needed for writing a robust and analysable specification. Hence, we decided to stick to a simple imperative structure with support for parametric polymorphism. Features like pointers, dynamic memory allocation and inheritance are too low level to be relevant at the specification level, and have been left out.

We note that the choice of some of the finer features in the action language part of StaBL were also governed by implementation issues. A more \emph{`StaBL-ised'} version of StaBL may be designed to integrate a lot more closely to one of more well-known programming languages.

Note that states in StaBL \emph{roughly} correspond to views or pages. Transitions correspond to navigation from one page/view to another. Also note that the variables in the specification may or may not correspond to program variables. For example, in our implementation \lstinline@Student.students@ translates to a database table, while \lstinline@Student.LoggedOut.user@ translates to an input text field in the \emph{LoggedOut} web-page. Specification variables may not even have any equivalent data in the resultant implementation.

\section{Semantic Aspects} \label{s:sem}
\subsection{StaBL Types}
\label{s:types}
StaBL supports following types: Basic types - int, boolean, string; Structures; Functions; Containers, e.g. lists, tuples, maps and sets;

Following points are to be noted about all the types in the language:
\begin{enumerate}
	\item All basic types and containers are \emph{native types}. All other types are \emph{user defined types}.
	\item All types are globally declared, whether native or user defined. That is, variables can be declared anywhere in the specification of all types, approximately like it is in C.
	\item Containers and functions are polymorphic. The variable declarations for such objects must substantiate at the time of declaration. For example:
	\lstset{style=mycode}
	\begin{lstlisting}[mathescape=true,frame=single, xrightmargin=.05\textwidth]
registeredUsers : map $\btl$ string, User $\btr$;
	\end{lstlisting}
\end{enumerate}

As mentioned above, all types are globally visible.

\subsubsection{Structures.}
A \lstinline@struct@ type declaration looks like as follows:
For example:
\lstset{style=mycode}
\begin{lstlisting}[mathescape=true, frame=single, xleftmargin=.05\textwidth, xrightmargin=.05\textwidth]
struct Duration = {
  startTime : int;
  endTime : int;
}
\end{lstlisting} 

As of now, we don't allow recursive types.

\subsubsection{Functions.}
Functions map a list of input types to an output type. For example: $sum : fun(int \times int \rightarrow int)$ declares a function $sum$ which takes two $int$s and returns an $int$. In our current implementation of StaBL compiler, we allow function declarations, but function definition is not implemented. The semantics of a set of library functions acting on our container types are defined, which are used in the symbolic simulation of StaBL specifications for static analysis purposes. Functions are not allowed to cause external side-effects, i.e., if a function contains an assignment to any variable which is not local to the function, it will not typecheck. However, assignments are allowed at the local level, i.e., the value returned from a function can be assigned to a variable at the call site.

\subsubsection{Polymorphic Types.}
A polymorphic type can be thought of as a function ($type\ list \rightarrow type$) which takes a list of types as parameters and returns a type. This type of polymorphism, specifically called \emph{parametric polymorphism} in programming languages terminology, is a key feature of StaBL which allows creation of specifications which are flexible and re-usable on the one hand, and type-safe on the other.

A polymorphic type encapsulates a type expression inside it. The inner type expression is allowed to be any non-polymorphic type expression where type variables may take the place of types. Here are a few examples of polymorphic type declarations allowed in StaBL:
\begin{itemize}
	\item Polymorphic structures:
		
\lstset{style=mycode}
\begin{lstlisting}[mathescape=true, frame=single, xrightmargin=.05\textwidth]
struct$\btl$A$\btr$ S$_1${ ... }
struct$\btl$A, B$\btr$ S$_2$ {
  s$_1$ : S$_1$$\btl$B$\btr$;
  s$_2$ : S$_1$$\btl$A$\btr$;
  s$_3$ : S$_1$$\btl$int$\btr$;
  s$_4$ : string;
}
\end{lstlisting}

	\item Polymorphic functions:
\lstset{style=mycode}
\begin{lstlisting}[mathescape=true, frame=single, xrightmargin=.05\textwidth]
f$\btl$A, B$\btr$(p$_1$ : B, p$_2$ : S$_1$$\btl$A$\btr$, p$_3$ : S$_1$$\btl$int$\btr$, p$_4$ : string) : S$_1$$\btl$S$_2$$\btl$A,B$\btr\btr$
\end{lstlisting}

\end{itemize}

\subsubsection{Substantiation of Polymorphic Types.}
When a variable is declared as a polymorphic, it is really a \emph{substantiation} of a polymorphic type. By substantiation, we mean getting a concrete type from a polymorphic type by supplying the required type arguments. For example:

\lstset{style=mycode} 
\begin{lstlisting}[mathescape=true, frame=single, xleftmargin=.05\textwidth, xrightmargin=.05\textwidth]
p : pair$\btl$ int, boolean $\btr$;
\end{lstlisting}
\subsubsection{Containers.} \label{s:cont}
Currently, we have provided containers like lists, sets and maps. All of them are polymorphic types. There are library functions declared which allow interacting with these containers. For instance, $put\_map$ allows adding/updating a key value pair to a map, $get\_map$ allows retrieving a value corresponding to a given key from a map. All these functions are side-effect free, i.e. a call to $put\_map$ does not modify the map passed as argument; rather it return a new map object. The decision to make functions side-effect free may have performance implications if translated directly into a running program. However, as StaBL statecharts are really specifications, their objective is analysis and not implementation. Side-effect free functions ease many difficult issues for static analysis.

\subsection{Terminology}
Now, we introduce some terminology and notations to help a succinct and consistent presentation of further concepts:

StaBL specifications are referred to as statecharts. The sets of all states and transitions in a statechart $SC$ are given by $S(SC)$ and $T(SC)$. Enclosing state of a state $s$ is given by $\mathcal{S}_s(s)$ and its entry and exit codes are given by $\mathcal{N}(s)$ and $\mathcal{X}(s)$ respectively. Enclosing state of a transition $t$ is given by $\mathcal{S}_t(t)$ and its trigger, guard and action are given by $\mathcal{T}$, $\mathcal{G}$ and $\mathcal{A}$ respectively. Source and destination states of a transition $t$ are given by $\mathbb{S}(t)$ and $\mathbb{D}(t)$ respectively. Given two states $s$ and $s'$, $s \prec_1 s' = true$ if $s' = \mathcal{S}_s(s)$. $s \prec_1 s' = false$ otherwise. Given two states $s$ and $s'$, $s \preceq s' = true$ if for some integer $n > 0$, $\exists s_1, s_2, ..., s_n$ such that $s = s_1$, $s_1 \prec_1 s_2$, ..., $s_{n - 1} \prec_1 s_n$, $s_n = s'$. $s$ is called a descendent state of $s'$. Likewise, a state $s$ is an ancestor of another $s'$, if $s' \preceq s$. This is abbreviated as $s \succeq s'$. Both relations $\preceq$ and $\succeq$ are reflexive, anti-symmetric and transitive. For an atomic state $s$, $atomic(s) = true$. For a non-atomic state $s$, $init(s)$ gives its initial sub-state.

\subsection{Scoping and Typing in StaBL} \label{s:env}
As mentioned earlier, the scoping rules for variables in StaBL are driven by two conflicting requirements:
\begin{enumerate}
	\item to restrict visibility of variables to within states to achieve modularity;
	\item to allow restricted data-flow between states.
\end{enumerate}

Data being defined in one view and being used in another is an essential feature of web apps. Without this feature, it would be impossible to specify interesting web applications. A simple example is when a user logs into a web application, she is taken to her dashboard possibly displaying many other details about her already stored in the database of the application. Hence, on transitioning from the login page to the dashboard, what gets displayed is dependent on the user ID that is entered in the login page. How do we achieve this effect in a Statechart specification? A simple alternative would be to use global variables. However, this has all the disadvantages of a global variable, and could become a bottleneck creating a large specification. In StaBL, we achieve inter-state visibility of local variables by defining the following scoping rules: local variables of a state are readable on its outgoing transitions, and are writable on its incoming transitions. The scoping rules of StaBL are summarised follows:
\begin{enumerate}
\item Visibility of a variable in any part of the specification are defined in terms of the predicates $R$ (read) and $W$ (write). Variables with $R$ access are allowed to appear in expressions -- in the transition guards, on the RHS of assignments, etc.; and those with $W$ access are allowed as l-values, viz. in LHS of assignments.
%\item Three derived predicates $RW$ (read and write), $RO$ (read-only) and $WO$ (write-only) are defined in terms of $R$ and $W$ as follows: At a given point in the specification, for a given variable $v$, $RW(v) \iff R(v) \wedge W(v)$, $RO(v) \iff R(v) \wedge \neg W(v)$, and , $WO(v) \iff \neg R(v) \wedge W(v)$
\item All local variables of a state have $RW$ (i.e. $R \wedge W$) access in itself and all its descendent states.
\item On a transition $t$, all local variables of the source state $\mathbb{S}(t)$ and its ancestor states have $R$ access, and all local variables of the destination state $\mathbb{D}(t)$ and its ancestor states have $W$ access. All local variables of $\mathcal{S}_t(t)$ and its ancestor states have $RW$ access. All local variables of $\mathbb{S}(t)$ and its ancestor states up to and excluding $\mathcal{S}_t(t)$ have $RO$ (i.e. $R \wedge \neg W$) access.  All local variables of $\mathbb{D}(t)$ and its ancestor states up to and excluding $\mathcal{S}_t(t)$ have $WO$  (i.e. $\neg R \wedge W$) access. Local variables of descendent states of $\mathbb{S}(t)$ and $\mathbb{D}(t)$ are inaccessible on $t$.
\end{enumerate}

The local variables in StaBL and their scoping rules are implemented in the StaBL typechecker. During typechecking of the specification, all names are looked up in a (type) environment. An environment can be visualised as a linked list of declaration lists, shown as $D^1 :: ... :: D^{n-1} :: D^n :: \phi$
where $D^1$, $D^2$, ... $D^n$ are declaration lists, $D^1$ being the declaration list of the closest surrounding state. $\phi$ represents an empty declaration list, Operator `$::$' represents the list construction operator.

A lookup of a variable $v$ in an environment $\sigma$, denoted by $\sigma[v]$, consists of searching for $v$'s declaration in all the declaration lists in $\sigma$ starting from the left end. $v$ gets bound with the first matching declaration in this order.

%Also, an environment $\sigma$ is defined for a state $s$ ($\sigma^s$) or a transition $t$ ($\sigma^t$). States and transitions are referred to as \emph{principle elements} for the rest of the discussion.

Two environment types are defined as follows: Read environment $\sigma_R$ is where a name which is \emph{used} in an expression (i.e. used as r-value) is looked up in. Write environment $\sigma_W$ is where a name which is \emph{defined} in an expression (i.e. used as l-value) is looked up in.The environments for the principle elements are defined as summarised in table~\ref{t:env}. 
 
\begin{table}
\begin{center}
\begin{tabular}{| l | c | c |}
\hline
	          & \textbf{State} $s$   & \textbf{Transition} $t$ \\
\hline
	$\sigma_R$ & $\sigma(s)$ & $\sigma(\mathbb{S}(t))$ \\
	$\sigma_W$ & $\sigma(s)$ & $\sigma(\mathbb{D}(t))$ \\
\hline
\end{tabular}
\end{center}
\caption{Environments for states and transitions}
\label{t:env}
\end{table}

\begin{myframe}{$t_{Assign}$}
\inferrule
{\sigma_W \vdash vname : T \\ \sigma_R \vdash e : T}
{\sigma_R, \sigma_W \vdash vname \gets e \longrightarrow (nil, OK)}
\end{myframe}

Shown above is an example of a typechecking rule: $t_{Assign}$, the rule for an assignment statement. As mentioned above, the LHS of the statement, $vname$, is looked up in $\sigma_W$ while the the RHS $e$ is typechecked in $\sigma_R$. If both typecheck to some type $T$, then the typechecking of the assignment statement succeeds, giving a $nil$ type and $OK$ to indicate success.

\subsection{Typechecking Polymorphic Types}
Type checking of a polymorphic structure type declaration and a polymorphic function declaration proceed somewhat similarly to each other. The fields (whether fields in a structure or parameters in a function declaration) are identical in their roles, from type checking perspective. The rules for typechecking a polymorphic type declaration are informally outlined as follows:
\subsubsection{Polymorphic Type Declarations.}
\begin{itemize}
	\item Each field should reference a known type.
	\item If a referenced type is a polymorphic type, then the number of type arguments should be equal  to the number of type parameters of the declared polymorphic type.
	\item Each type should get bound to its declared type. In case of polymorphic type, the type argument should get bound to a known type or a type variable corresponding to the respective type parameter of the enclosing type, or a known polymorphic type substantiated appropriately with known types or known type parameters, recursively.
\end{itemize}

\subsubsection{Variable Instantiation with Polymorphic Types.}
When variables are declared using polymorphic types, they must be completely substantiated, i.e. they may not have any type variables as arguments to the declared type. The type must supply an appropriate number of type arguments.

\subsubsection{Variables of Polymorphic Types in Code.}
In code fragments involving reconciliation between types, e.g. equality or assignment, the type checking is done by doing structural equality test between types.
\begin{figure}
\begin{center}
\lstset{style=mycode}

\begin{tabular}{| c | c |}
\hline

\begin{lstlisting}[mathescape=true]
struct $\btl$A, B$\btr$ S{ ...}

x : S$\btl$int, int$\btr$;
y : S$\btl$int, int$\btr$;

y := x;
\end{lstlisting}
&
\begin{lstlisting}[mathescape=true]
struct $\btl$A, B$\btr$ S{ ...}

x : S$\btl$int, int$\btr$;
y : S$\btl$string, string$\btr$;

y := x;
\end{lstlisting} \\
\hline
(a) & (b) \\
\hline
\end{tabular}

\end{center}
\caption{Using variables of polymorphic types: (a) Correct usage; (b) Incorrect usage.}
\label{f:ptu}
\end{figure}

\vspace{0.3cm}
Fig.~\ref{f:ptu} shows a sample code of how variables of polymorphic types can be used in expressions and instructions. In the left sub-figure, \lstinline@x@ and \lstinline@y@ are of structurally identical types, i.e. \lstinline[mathescape=true]@S $\btl$int, int $\btr$@. This code will typecheck. On the other hand, the code fragment in the right sub-figure will not type check as the type arguments cause the polymorphic structure \lstinline@S@ to substantiate to two dissimilar types for \lstinline@x@ and \lstinline@y@.

\section{Related Works}
\label{s:rw}
There are various formalisms to model web applications, these are used to models and perform verification and testing on web applications - an extensive analysis on this is presented in \cite{doi:10.1002/stvr.401}. Here we analyse the works that use state transition diagrams to model web application. Statecharts, state transition diagrams, finite state automata are widely used to model systems/applications in various domains like hardware systems, embedded systems\cite{10.1007/978-3-540-24721-0_17}, softwares \cite{708566}, web applications \cite{883017,DEUTSCH2007442,884689,10.1007/978-3-540-39929-2_5} , websites\cite{deOliveira:2001:SMH:366836.366869,Andrews2005,FLORES2008103}, GUI \cite{Nikolai05modelingand}. The necessity to use formal methods to model requirements is emphasised in \cite{10.1007/978-3-319-16101-3_11,10.1007/978-3-319-16101-3_12,10.1007/978-3-642-02050-6_8}. Li\cite{10.1007/978-3-319-16101-3_12} takes the natural language requirements and defines a methodology to convert into $TeAL$ language specifications, that are temporal in nature in order to detect inconsistencies and Perrouin\cite{10.1007/978-3-642-02050-6_8} provides a methodology to combine various models used in software development by providing a meta model and then detect inconsistencies in the combined model. These works are in specific to the method to convert Natural Language Requirements to certain formalism and in specific, does not to handle the features of web applications. In the works that use statechart-like models for web applications - Miao\cite{4276301} models external observable behaviour of the system using a Kripke structure with states representing a Page or request and atomic propositions representing the requests/triggers that are enabled in a state. This work considers dynamically generated pages, ActiveX, java beans, JSP, ASP, sub pages(a frame inside a page) etc. but the model used for generating properties (the object relationship diagram), doesn't consider the "result of data processed". Thus, a property generated from it  "StudentView should only be reached after Login" will not ensure whether $StudentView$ can be reached when the credentials are false. In Han's \cite{Han:2006:MVA:1145581.1145645} work, such constraints are termed as adaptive navigation. A FarNav approach is proposed using SMV and CTL to check for various problems like security, dead end, unreachability, reach a page through certain number of steps, allowed pages etc. However, the this work doesn't use action language, hierarchy and pseudo-states.  De Oliveira\cite{deOliveira:2001:SMH:366836.366869} presents a work to model hypermedia applications. Statechart's features like hierarchy and parallelism is used for specifying display characteristics, browsing semantics, parallel interactions of hypermedia - action language, data-flow has not been used in this model. StateWebCharts\cite{10.1007/978-3-540-39929-2_5} - a formalism claimed to have been developed specifically for modelling web applications. An extension of classical statechart formalism proposes new type of states - static, transient, dynamic and external. Combination of events(user/system-generated) determines the next state in the statechart rather than a single event. External states are used to model those entities that are external(like 3rd party modules, external websites etc.) and Dynamic state can be used to denote new transitions that can be generated at run time which does not exist/cannot be determined during design. However,in our work, we would not want such feature as it would lead to incomplete specifications and also in future work, we aim to use static verification approaches to identify the requirement bugs. Alpuente\cite{ALPUENTE201479} provides a rewriting logic approach to specify web applications - this work handles specifying the DB interactions, session interactions, server-client interactions for each operation, these are implementation aspects which are abstracted out in our work handles formal specifications at the requirements phase. 

\section{Conclusion and Future work}
\label{s:conc}
A lot of research exists in the area of formal verification of Statecharts and in using Statecharts for modelling and verifying specific aspects of web applications. But a gap exists in the use of Statecharts as a general purpose functional specification language for web apps. This paper is our attempt to address this gap. We have demonstrated that Statecharts features need to be tuned to allow scalable development of specifications. We have illustrated one such feature, namely statically scoped variables and inter-state data-flow. On the method side, we have demonstrated that a tighter integration of the action language with the Statechart model, added with rich features of modern programming languages (e.g. static typing with parametric polymorphism) gives greater power in the hands of the software engineer. Above features have been implemented in StaBL language, for which we have developed a compiler and have developed a number of case-studies.

One of the key advantages of formal specification is the possibility of early detection of specification issues using automated means. We have developed a novel approach to do formal verification of StaBL specifications which we have used to detect several specification issues in our case-studies. We have kept the presentation of our verification approach reserved for a separate paper. 

Our future work focuses on strengthening our formal verification approach and conducting experiments on larger case-studies. Our plans also include exploring automated test generation and UI synthesis from StaBL specifications.
\bibliographystyle{plain}
\bibliography{references}\end{document}